\documentclass[preprint]{aastex}
\shorttitle{Star Formation History of NGC 6822}
\shortauthors{Wyder}
\begin{document}
\title{The Star Formation Histories of Four Fields Spanning the Minor Axis
of NGC 6822\footnote{Based on observations made with the NASA/ESA 
{\it Hubble Space Telescope}, obtained at the Space Telescope Science
Institute, which is operated by the Association of Universities for
Research in Astronomy, Inc., under NASA contract NAS 5-26555. These
observations are associated with proposal 8314.}}
\author{Ted K. Wyder}
\affil{California Institute of Technology, 1200 E California Blvd, MC 405-47,
Pasadena, CA 91125}
\email{wyder@srl.caltech.edu}

\begin{abstract}

The star formation histories of four fields within the Local Group
dwarf irregular galaxy NGC 6822 are presented. Each of the fields was
imaged by the WFPC2 aboard the {\it Hubble Space Telescope} and were
used to obtain $VI$ color-magnitude diagrams for each field reaching
$V\simeq26$.  The magnitude of the tip of the red giant branch and the
red clump were used to determine distances to NGC 6822 that are
consistent with previous ground-based measurements. The distance,
extinction and star formation history were also determined by fitting the
entire color-magnitude diagram in each field. The distances from these
fits are consistent with the other determinations within the estimated
errors once the systematic effects of uncertainties in the
age-metallicity relation are taken into account. The extinction varies
among the four fields from approximately the foreground Galactic value
to $\approx 0.4$ mag higher in $V$ and roughly correlates with the
$60\micron$ surface brightness. The star formation histories in the
four fields are similar for ages $\gtrsim 1$ Gyr and are relatively
constant or somewhat increasing with time. These old star formation
rates are comparable to that expected from the typical gas surface
densities at these galactocentric radii and suggest that no large
scale redistribution of gas or stars is required to account for the
inferred star formation rates.  Three of the fields show a drop of a factor of
$\sim 2-4$ in the star formation rate about 600 Myr ago while the
remaining field centered on the bar shows an increase.

\end{abstract}


\section{Introduction}

Much progress has been made in understanding the evolution of galaxies
both from comparing the integrated properties of galaxies at various
redshifts and from detailed observations of individual nearby
galaxies. In particular, the dwarf galaxies in the Local Group provide
a unique opportunity to study the evolution of individual systems.  Due
to their close distances,  detailed information about their star
formation histories (SFHs) can be derived from the color-magnitude
diagrams of their resolved stellar populations. With the exceptions of
the Magellanic Clouds, the satellite galaxies of the two most massive
galaxies in the Local Group, M31 and the Milky Way, are predominantly
dwarf spheroidals and dwarf ellipticals which have little current star
formation and whose evolution has presumably been influenced by their
proximity to a more massive galaxy. The remaining relatively isolated
galaxies in the Local Group are predominantly gas-rich dwarf irregular
(dIrr) galaxies which give us the opportunity to study the evolution of
isolated galaxies.

The available evidence indicates that no two dIrrs share the exact
same SFH although nearly all of them show evidence for a significant
population of intermediate and old stars (ages $> 1$ Gyr) in addition
to their ongoing star formation.  Furthermore, for many dIrrs such as
WLM \citep{minniti97}, Phoenix \citep{martinez99} and Leo A
\citep{dolphin02a}, the young blue stars are concentrated in the
central regions while the outer regions contain a predominately older,
redder stellar population.

NGC 6822 in particular is a Local Group dIrr whose structure fits the
patterns described above for dIrrs. This galaxy consists of a bar
oriented roughly North-South that contains most of the current sites
of star formation which is surrounded by a lower surface brightness
area containing a higher proportion of redder, fainter stars
\citep{hodge77, hodge91}. Based upon wide-field $BRI$ images from the
8m Subaru Telescope reaching down to a level of $M_B \approx 0$,
\citet{komiyama02} found that the red stars are distributed
symmetrically about the center of the galaxy. While the brightest blue
stars are concentrated in the central bar, main sequence stars were
detected in the bar and in two arms on either side 
stretching $\sim40\arcmin$ from the
South-East to the North-West. The blue star distribution
follows fairly closely the distribution of HI gas
\citep{deblok00,weldrake02}.

The SFH of NGC 6822 has been investigated most recently by
\citet[b]{gallart96a} from ground-based data and by \citet{wyder01}
from {\it HST} data. Both studies found that the observed CMDs are
best fit with star formation beginning $12-15$ Gyr ago although
younger ages are allowed if a high initial metallicity is
assumed. Based upon their data covering the central $11.2\arcmin
\times 10.4\arcmin$ area of NGC 6822, \citet{gallart96b} found
evidence for an increase in the star formation rate
(SFR) $100-200$ Myr ago over the entire
area surveyed with the largest increases seen at the ends of the bar.
\citet{hodge80} found a similar peak in the cluster age distribution 
$75-100$ Myr ago. In \citet{wyder01}, the SFH of NGC 6822 was
analyzed in three fields lying in the bar and two other fields sampling
more of the outer regions. The overall shapes
of SFHs among the five fields were similar and showed a more or less
constant or perhaps somewhat increasing SFR with time. However, for
ages less than about 600 Myr there were significant variations,
implying that stars formed within this time are not yet very well
mixed throughout the galaxy. As in the ground-based studies,
the bar fields showed an increase in the recent SFR while the SFR
decreased in the outer regions.

This paper further explores the SFH of NGC 6822 begun in
\citet{wyder01} based upon CMDs of four fields that span the minor
axis of the bar of NGC 6822, one of which was included in
\citet{wyder01}. The main goal of this work is to generate a SFH
profile of the minor axis of NGC 6822 in order to compare the
differences between the SFH of the bar with the outer regions. 
It remains unclear whether the stars have always formed in
the central regions and then migrated outwards or whether the stars we
observe today in the outer regions were actually formed there as
well. A first step in trying to understand the origin of these stellar
populations differences is to understand the differences in the ratio
of young to old stars between the central and outer regions.   A
comparison of the star formation rate as a function of time in
different regions within each galaxy with the distribution of atomic
and molecular gas will provide additional clues to the evolution of
these systems.

This paper is organized as follows. \S{2} presents the data and 
photometry while \S{3} briefly summarizes the analysis methods
used to extract information about the SFH from the data. The results
are presented in \S{4}. 

\section{Observations and Photometry}

\begin{figure}
\plotone{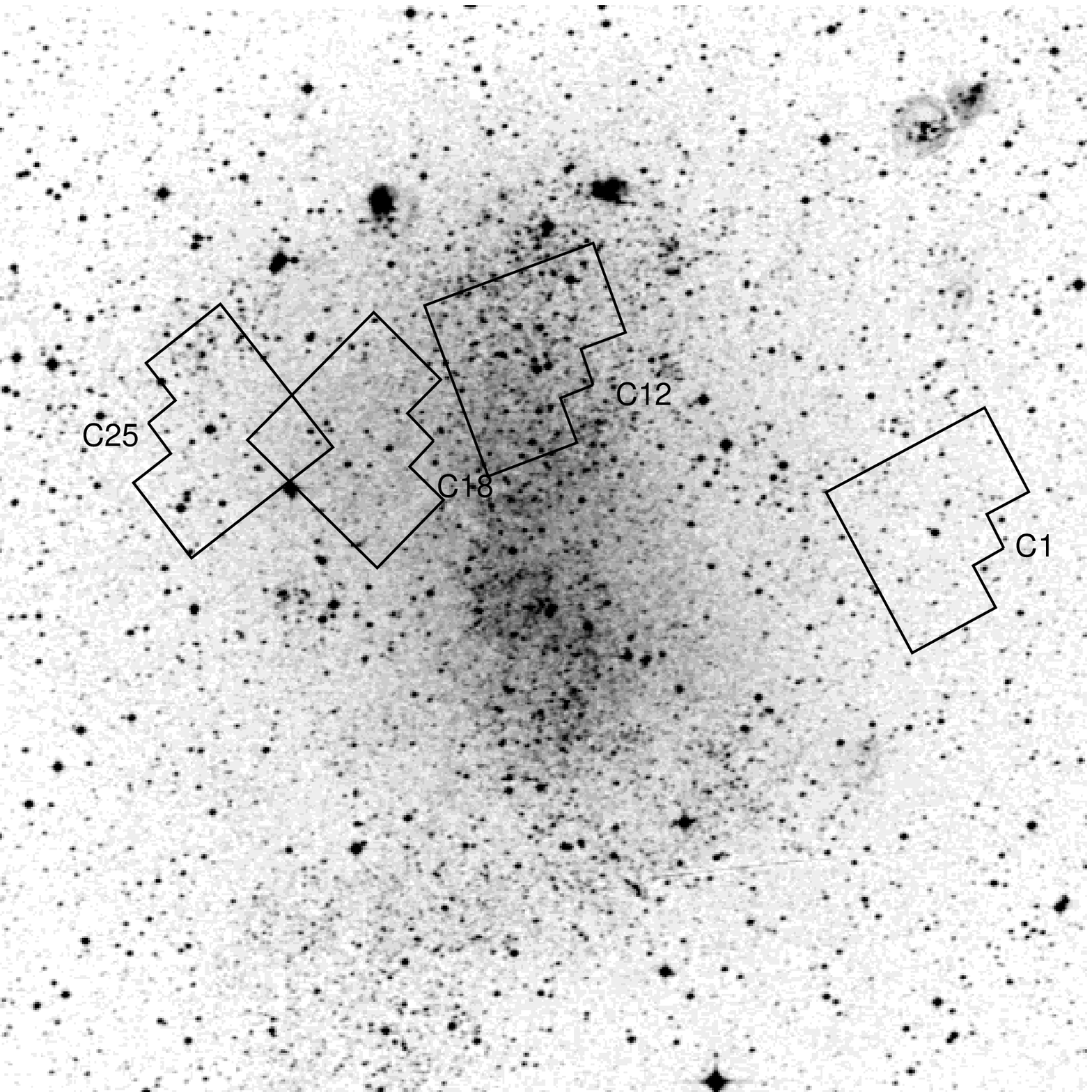}
\caption{Digitized Sky Survey image of NGC 6822 (North up, East to the left).
The field-of-view is $15\arcmin$, or about the same size the galaxy's
$D_{25}$ diameter.  The positions of the four WFPC2 fields analyzed
here are shown. Each is labeled by the star cluster from the list in
\citet{hodge77} that is centered on the PC chip.\label{n6822dss}}
\end{figure}

The data presented here consist of images taken with the WFPC2 aboard
{\it HST} through the $F555W$ and $F814W$ filters,
approximately corresponding to the ground-based $V$ and $I$ filters,
respectively.  In each field the PC chip was centered on one of the open
clusters C1, C12, C18 and C25 from the list in \citet{hodge77} and
these field centers were selected to span the bar's minor axis.  The
locations of these four fields are shown on an
Digitized Sky Survey image of NGC 6822 in Figure
\ref{n6822dss}. Two exposures per filter at each pointing were obtained
in order to facilitate the removal of cosmic rays.
A list of the files analyzed in this paper is shown in Table \ref{obs_log}.
The data on the C25 field were included in the data presented 
previously in \citet{wyder01}. The clusters appearing in each of these
fields have been analyzed by \citet{seth03}.

I obtained photometry for the stars in these images using version 1.1
of the PSF-fitting software package HSTPHOT \citep{dolphin00} which I
obtained from the author's web
site.\footnote{http://www.noao.edu/staff/dolphin/hstphot/} 
Following the prescription described in more detail in \citet{dolphin00}
and the HSTPHOT manual, bad pixels in each image were
masked out and the two images per filter of each field were added
together while at the same time detecting and masking out pixels
likely affected by cosmic rays.  HSTPHOT was run on the data presented
here in a nearly identical manner to the analysis described in
\citet{wyder01} and the reader is referred to that paper for a more
detailed description of the analysis as well as a more detailed
discussion of the photometric errors and completeness.  The only
difference here is that a stricter $\chi$ limit of $\chi < 2.5$ was
imposed. HSTPHOT automatically determines an aperture correction for
each field and chip by comparing the PSF-fit magnitudes with
magnitudes measured within a $0.05\arcsec$ radius aperture. The
standard deviations of the aperture corrections are $\approx 0.05$
magnitudes and represent the dominant source of error in the
photometry zeropoint.  As in \citet{wyder01}, artificial star tests
were conducted for each of the fields and the results were used to
describe the photometric errors and completeness in the analysis
presented in the following section. Furthermore, only stars detected
on the three WF chips were included in the final photometry list due
to the different completeness and errors for stars detected on the PC
chip.  There are a total of 21,851, 50,004, 47,614, and 37,001 stars
detected in the C1, C12, C18 and C25 fields, respectively.

\begin{figure}
\plotone{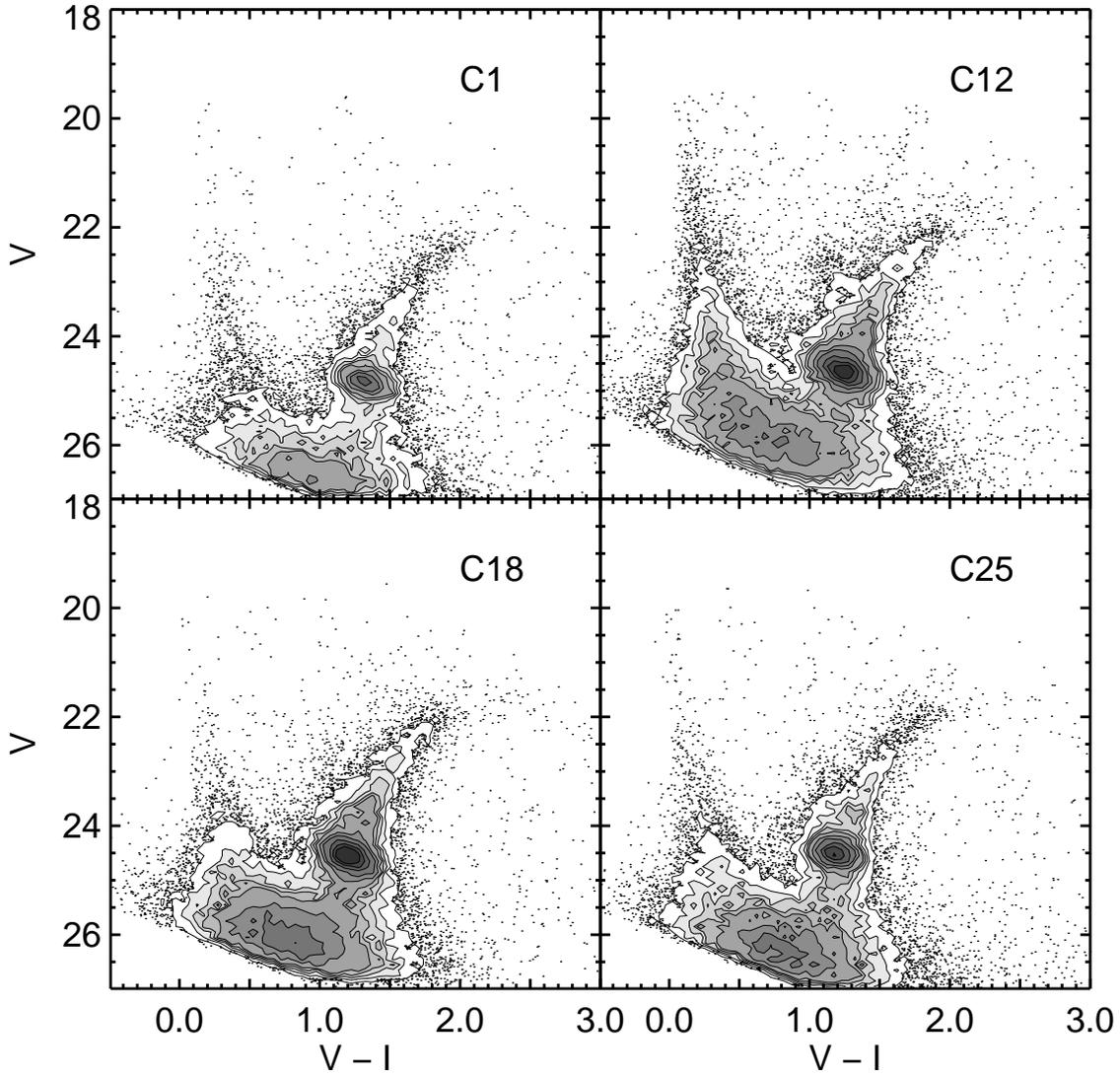}
\caption{Observed $VI$ CMDs for all stars detected in the WF chips of
each field. The contour levels correspond to stellar densities of 20,
40, 60, 80, 100, 150, 200, 300, 400 and 500 stars decimag$^{-2}$.  For
those areas of each diagram with a density of stars smaller than the
lowest contour, each individual star is plotted.\label{cmds}}
\end{figure}

The resulting $VI$ CMDs for each field are  shown in Figure \ref{cmds}
and reach levels of $V\simeq26$ for the bluest stars and somewhat
fainter for the redder stars.  All of the fields contain stars with a
wide variety of ages from young main sequence stars a few 10s of Myr
old up to red giant branch stars many Gyr old. The relative strength
of the main sequence relative to the red clump and red giant branch is
higher in the central C12 field compared to the other three fields
that sample less of the bar and more of the outer regions. Note also
the difference in color of the main sequence which becomes bluer going
from the C1 to the C25 field (or equivalently, from West to East
across the bar). This changing color of the main sequence is reflected
in the varying value of the extinction from the CMD-fitting described
in the next section. The most prominent feature in all  of the
diagrams is the red clump appearing at $V\approx25$ and
$1.0<(V-I)<1.5$. Stars in the red clump are core-Helium burning
stars with ages of $1-10$ Gyr and the prominence of the red clump
indicates the strength of the
intermediate age stellar populations in NGC 6822.  As is evident from
the CMDs, the magnitude and color of the red clump decrease going from
the C1 to the C25 fields which provides additional evidence for
variation in dust extinction among the four fields.

\section{Analysis}

In this section I first determine the distance from the CMDs of
each of the fields using the tip of the red giant branch (TRGB) and
the magnitude of the red clump and then proceed to summarize the methods
used to infer the star formation histories from the data. The red clump
and TRGB distances provide a useful external check on the distances
determined from the fits to the full CMDs.

\subsection{Distance from the TRGB magnitude}

\begin{figure}
\plotone{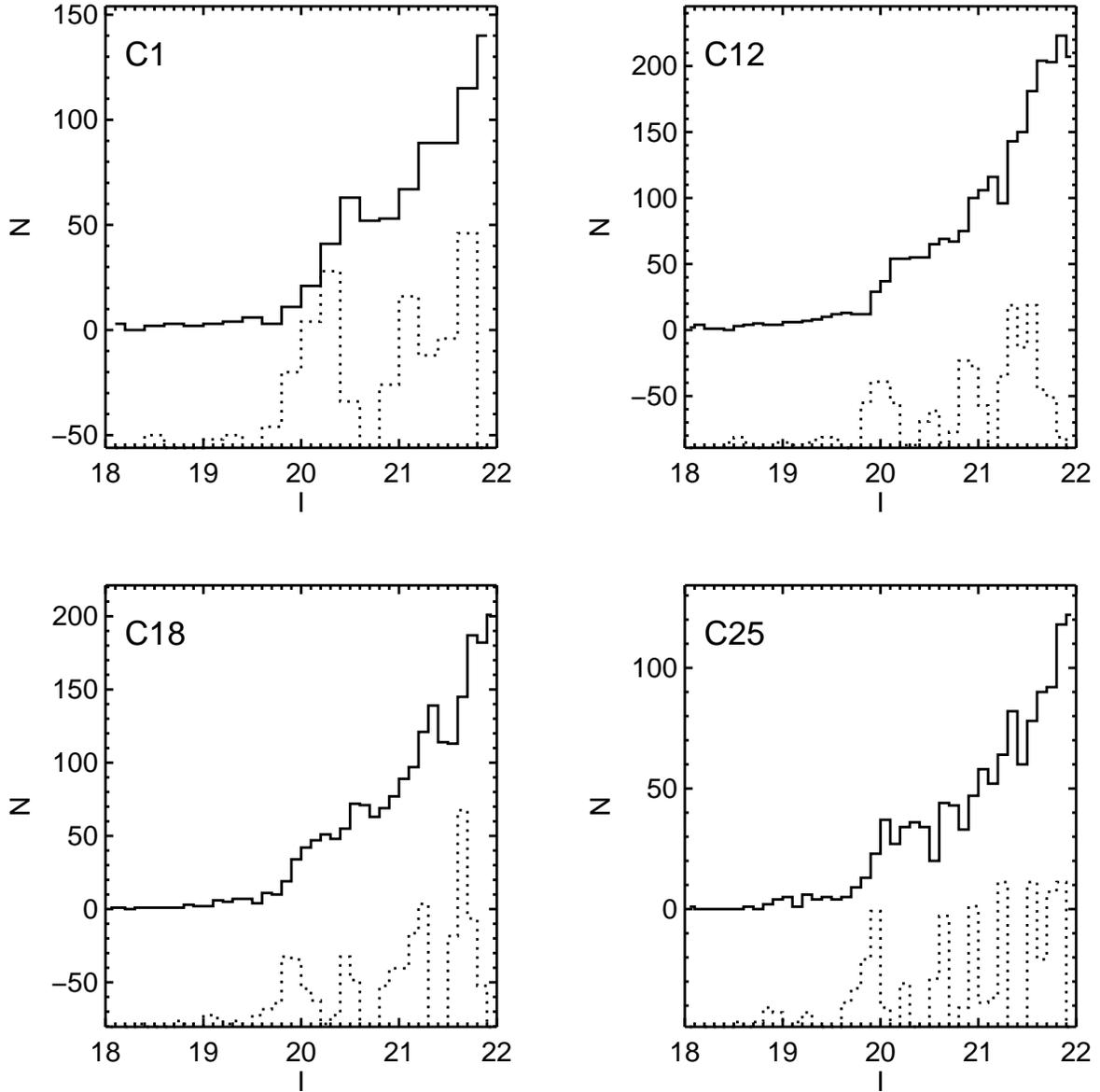}
\caption{$I$-band RGB magnitude distribution for each of the fields. In each
panel the magnitude distribution for stars with $1.0<V-I<2.2$ is plotted as
the solid line. The dotted line is the magnitude distribution convolved with
an edge-detecting Sobel kernel (-2,0,2) and has been offset for clarity
by an arbitrary amount in each panel. No correction for extinction has been
applied.\label{rgblf}}
\end{figure}

The TRGB has been shown to be a reliable distance
indicator for galaxies with ${\rm [Fe/H]}<-0.7$ \citep{lee93}. The
$I$-band magnitude distributions for stars with $1.0<(V-I)<2.2$ for
each field are plotted in Figure \ref{rgblf} as the solid
lines. Following the analysis of \citet{lee93}, these distributions
were convolved with an edge-detecting Sobel Kernel of $(-2,0,2)$ to
produce the dotted line in each plot. The position of the first peak
in the convolved functions in each panel was used to determine the tip
magnitude while the widths of the peaks were used to estimate the
error. The resulting $I$-band magnitudes of the RGB tip are listed in
Table \ref{rgbtab}.

I have calculated the TRGB distance for each of the four fields using
the prescription described in detail in \citet{lee93}. This method
calculates the distance modulus from the following equation:
\begin{equation}
(m-M)_0 = I_{TRGB} - A_I + BC_I - M_{bol,TRGB}
\end{equation}
where $I_{TRGB}$ is the $I$ magnitude of the TRGB, $A_I$ the $I$-band
extinction, $BC_I$ the bolometric correction and $M_{bol,TRGB}$ the bolometric
magnitude at the TRGB.
The extinction for each field
was taken from the $A_V$ values listed in Table \ref{besttab} and converted
to $A_I$ using the Galactic extinction law. The value of $BC_I$ for
each field was calculated from the $(V-I)_{TRGB}$, the color
at the TRGB, using the relation 
\begin{equation}
BC_I=0.881-0.243(V-I)_{TRGB}
\end{equation}
determined by \citet{dacosta90} from Milky Way globular cluster
RGBs. \citet{dacosta90} also give $M_{bol,TRGB}$ as a function
of [Fe/H]: 
\begin{equation}
M_{bol,TRGB}=-0.19{\rm [Fe/H]} - 3.81.
\end{equation}
Finally, the metallicity was estimated from the relation given
in \citet{lee93}:
\begin{equation}
{\rm [Fe/H]} =-12.64+12.6(V-I)_{-3.5}-3.3(V-I)_{-3.5}^2
\end{equation}
where $(V-I)_{-3.5}$ is the color of the RGB at $M_I=-3.5$, or $\approx 0.5$
mag below the TRGB. The values of $(V-I)_{TRGB}$ and $(V-I)_{-3.5}$
were determined as the median color of RGB stars with $I$-band
magnitudes within $\pm0.1$ mag at the tip or 0.5 mag below.
The resulting values of $(m-M)_0$ are
listed in column (5) of Table \ref{rgbtab} and are consistent to within
the errors with the ground-based TRGB distance
modulus of $(m-M)_0=23.4\pm0.1$ as well as the Cepheid value of 
$(m-M)_0=23.49\pm0.08$ \citep{gallart96c}.

\subsection{Distance from the red clump magnitude}

In the past few years, there has been interest in the possible use of
the red clump as a distance indicator. Using parallaxes from the {\it
Hipparcos} catalog for nearby Galactic red clump stars,
\citet{paczynski98} determined the absolute magnitude of the red clump
in the solar neighborhood to be $M_I=-0.23\pm0.03$.  This calibration
was then used to determine the distance to the Galactic bulge
\citep{paczynski98}, M31 \citep{stanek98a} and the Magellanic Clouds
\citep{udalski98,stanek98b}. In the case of M31, the red clump
distance agreed to within the uncertainties with the Cepheid distance
while the red clump distance moduli for the Magellanic Clouds
were found to be $\sim 0.5$ mag
smaller. \citet{cole98} argued that $\approx0.3$ mag of this
difference is due to the different age and metallicity distribution of red
clump stars in the Magellanic Clouds as compared to those in the solar
vicinity. From observations of the red clump magnitude in a sample of
Galactic open clusters with metallicities in the range $-0.4\lesssim
{\rm [Fe/H]} \lesssim 0.2$ and ages between $\sim 2$ and $\sim 9$ Gyr,
\citet{sarajedini99} found that the variation in red clump magnitude
with metallicity and age largely agreed with that expected from the
\citet{bertelli94} isochrones. \citet{girardi01} investigated in more
detail the theoretical dependence of the red clump absolute magnitude
on age and metallicity using the \citet{girardi00} isochrones and
stressed the need to account for this dependence when attempting to
use the red clump as a distance indicator.  Furthermore, their models
confirmed that red clump distance moduli to the Magellanic Clouds
should be $\sim0.2-0.3$ mag larger once these population effects are
included.

\begin{figure}
\plotone{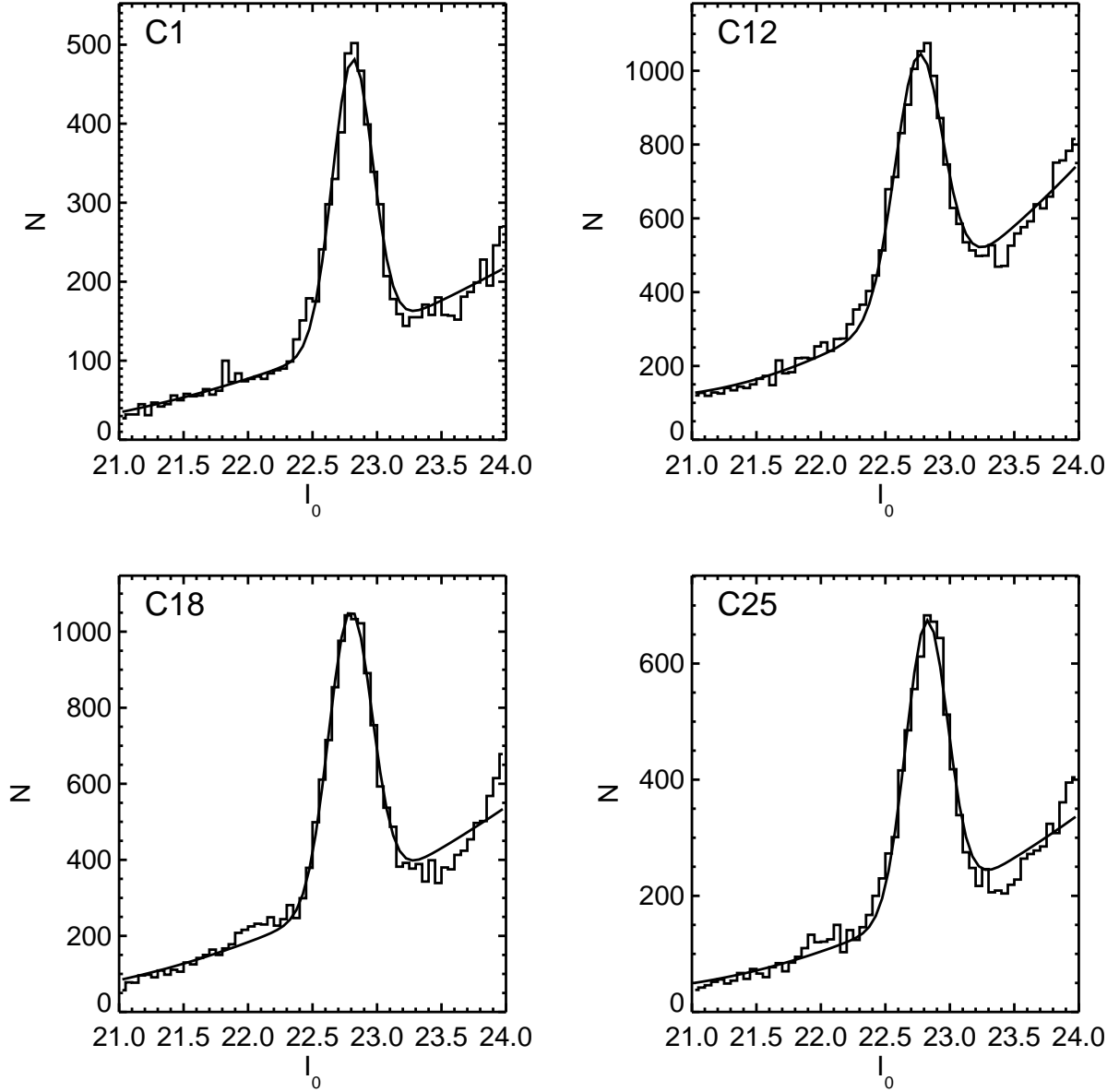}
\caption{$I$-band magnitude distribution for each field
in the region of the red clump. Each panel plots the magnitude distribution
for stars with $0.5<(V-I)_0<1.5$ as a histogram. The smooth curve
in each panel is the best-fit Gaussian plus quadratic to the data. In each
field, the data were corrected for extinction using the values determined
from the CMD-fitting, as described in the text.\label{rclf}}
\end{figure}

I have
estimated the distance to NGC 6822 based upon the apparent magnitude
of the red clump in conjunction with the theoretical results of
\citet{girardi01}.
In Figure \ref{rclf} the $I$-band magnitude distribution in the
vicinity of the red clump for each field is shown for stars with $0.5<
(V-I)_0 <1.1$, where each of the distributions has been corrected for
extinction using the $A_V$ values from the SFH solutions listed in
Table \ref{besttab}.  Following \citet{paczynski98}, I have fit each
of the distributions in Figure \ref{rclf}  with a function that is the
sum of a 2nd order polynomial and a Gaussian. This function is given by
\begin{equation}
n(I_0) = a + b(I_0-I_{0,m}) + c(I_0-I_{0,m})^2 +
\frac{N_{RC,I}}{\sigma_{RC,I} \sqrt{2\pi}} 
\exp{\left[ \frac{-(I_0-I_{0,m})^2}{2\sigma_{RC,I}^2} \right]}
\end{equation}
where $I_{0,m}$ represents the magnitude where the red clump peaks.
An analogous
function was fit also to the $V$-band magnitude distribution in the
region of the red clump except that the quadratic term was set to
zero.  The results of the fits for both $V$ and $I$ are listed in
Table \ref{rclftab}.  The average values of the red clump magnitude
among the four fields
are $V_{0,m}=23.61$ and $I_{0,m}=22.80$.

In order to derive distances from these apparent magnitudes,  it is
necessary to assume a value for the absolute magnitude of the red
clump. \citet{girardi01} give tables listing the variation of the red
clump magnitude with age and metallicity for stars formed in an
instantaneous star formation burst. As explained in detail in
\citet{girardi01}, these tables can be used to predict the absolute
magnitude of the red clump for any arbitrary SFH. In order to
understand the uncertainties due to the unknown metallicities of the
clump stars, I have calculated the absolute magnitude of the red clump
assuming a constant metallicity with two different values of ${\rm
[Fe/H]}$, namely ${\rm[Fe/H]}=-0.7$ and ${\rm [Fe/H]}=-1.7$.  The
higher metallicity is similar to the highest values for the current
metallicity of NGC 6822 \citep{pagel80,skillman89} while the lower
value is the lowest metallicity included in the \citet{girardi01}
calculations. Using their tables in conjunction with equation (6) of
their paper, I have calculated red clump $V$ and $I$-band absolute
magnitudes for each of the fields. In each field, I assume the SFR(t)
from the results of the CMD-fitting plotted in Figure \ref{sfh1}.
Since the SFR results themselves depend upon the distance to NGC
6822, this reasoning is somewhat circular. However, the shape of the
SFHs is not strongly dependent on the distance and using simply a
constant SFR(t) would yield similar results for the red clump absolute
magnitude. The resulting average values for $M_I$ over the four fields
are $M_I=-0.44$ and $M_I=-0.64$ for the high and low metallicity
cases, respectively. Similarly, I obtain values of $M_V=+0.44$ and
$M_V=-0.08$ for the high and low metallicity cases, respectively.

Applying these values to the average of the red clump magnitudes
listed in Table \ref{rclftab}, I find distance moduli of
$(m-M)_0=23.36\pm0.18$ for the $V$-band and $(m-M)_0=23.34\pm0.10$
for the $I$-band. The values of $(m-M)_0$ were calculated by averaging
the red clump absolute magnitudes for the high and low metallicity
cases while the uncertainty is simply one-half the difference.
While slightly lower than the TRGB and Cepheid distance moduli
determined by \citet{gallart96c} and in \S{3.1} of this paper,
the values for the red clump distance
agree to within the uncertainties with the other determinations.

\subsection{Star formation history analysis}

The analysis of the star formation histories in the four fields
presented here relies
upon the methods developed by \citet{dolphin97,dolphin02} which I have
implemented in a set of IDL routines. The analysis is nearly the
same as in \citet{wyder01} and thus only the differences with that
paper will be emphasized here.

First a set of stellar evolutionary isochrones must be selected to
generate the model Hess diagrams. I rely upon the same set
of isochrones as in \citet{wyder01}.  This primarily consists of the
isochrones presented in \citet{girardi00} which describe the evolution
of stars with masses from 0.15 to 7 $M_{\odot}$ for metallicities
ranging from $Z=0.0004$ to $Z=0.03$. The $Z=0.0001$ isochrones from
\citet{girardi96} were added to extend the models to lower
metallicities while stars more massive than 7 $M_{\odot}$ were included
using the older \citet{bertelli94} isochrones. A set of files
containing the  combined set of isochrones was kindly provided by
Andrew Dolphin.

The CMD analysis methods of \citet{dolphin97,dolphin02} fit the
observed data in the form of a binned CMD, or Hess diagram. A series
of time bins are also chosen for the SFH solution.
Assuming a distance,
extinction, metal enrichment history and initial mass function (IMF),
a Hess diagram is calculated for each time bin assuming a constant SFR
within that time bin and none at other times. The errors and completeness
in the photometry are included in the models using the results of the
artificial star tests. 
The linear combination
of these ``basis'' CMDs that best reproduces the observations
determines the SFR(t). The model and observed diagrams are compared
using a fit parameter derived from the Poisson distribution that is
the Poisson equivalent of the $\chi^2$ parameter derived from the
Gaussian distribution.  Starting from the Poisson distribution,
\citet{dolphin02} derived the following expression for the fit
parameter:
\begin{equation}
\Upsilon = 2 \sum{m_i} - n_i + n_i \ln{\frac{n_i}{m_i}}
\end{equation}
where $m_i$ is the number of model stars in the $i$th bin of the
model Hess diagram and $n_i$ is the number of observed stars in the
same bin.  The SFH which minimizes the value of $\Upsilon$ is the SFH
which best reproduces the observations \citep{dolphin02}. The average
and variance of $\Upsilon$ in each bin is one and two, respectively.  The
average and variance of $\Upsilon$ summed over the entire Hess
diagram depends
upon the number of observed stars as well as the number of Hess
diagram bins contributing to the solution \citep{dolphin02}.

In the equation for the fit parameter $\Upsilon$ in \citet{dolphin02},
a term is included in the models to account for the number of foreground
Galactic stars. \citet{ratnatunga85}
have predicted the number counts of Galactic stars in the direction
of several Local Group galaxies, including NGC 6822. According to their
results, a total of 270 Galactic stars with $17<V<27$ are predicted
to lie within each WFPC2 field-of-view of which $\approx 80\%$
have colors $(B-V)>1.3$. This level of contamination is small
compared to the $>10^4$ stars present in each of the CMDs and thus
the foreground contamination has been neglected in all of the model
Hess diagrams.

All of the SFH solutions presented in this paper were fit using
0.1 mag wide bins in $V$ and 0.05 mag wide bins in $(V-I)$.
A set
of nine time bins were also chosen for the SFH solution that span
ages from 0.01 to 15 Gyr ago.

One of the most important
ingredients used in calculating the model Hess diagrams
is the metal enrichment history.
For the oldest stars, some measurements
of their metallicities are beginning to be made. \citet{cohen98}
measured a metallicity of ${\rm[Fe/H]}=-1.95\pm0.15$ and age of
$11_{-3}^{+4}$ Gyr for the star cluster Hubble VII based upon its
integrated spectrum. For the field stars, \citet{tolstoy01}
measured metallicities for 23 red giant branch stars based upon
the equivalent widths of the \ion{Ca}{2} triplet near 8500 \AA.
Their distribution peaks at ${\rm [Fe/H]} \approx -1.0$ with a tail of stars
reaching ${\rm [Fe/H]} \approx -2.0$.

The current metallicity in NGC 6822 has been determined both from
spectra of its \ion{H}{2} regions as well as from spectra of individual
massive stars. For the \ion{H}{2} regions, \citet{pagel80} and
\citet{skillman89} measured an oxygen abundance of $12 + {\rm
log}(O/H)=8.25\pm0.07$, corresponding to ${\rm [Fe/H]}=-0.7$ for a
solar O/Fe ratio. For a sample of six \ion{H}{2} regions within NGC
6822, \citet{chandar00} measured a value of $12 + {\rm
log}(O/H)=7.91\pm0.06$ (${\rm [Fe/H]}=-1.0$).

The precise form of the age-metallicity relation in NGC 6822
is unknown. In fact, in the presence of significant outflows or inflows
of gas, the metallicity may not necessarily increase monotonically
with time. However, the sparse evidence summarized above indicates that
the current metallicity of the gas in NGC 6822 is larger than that measured
for the oldest star cluster Hubble VII although there is little
observational constraints at intermediate ages.

In principle with very good, deep photometry,
it is possible to simultaneously
determine the SFR(t) and the metal enrichment history from the Hess
diagram fits. However, there are significant degeneracies between age
and metallicity in the shape and placement of the red clump and 
red giant branch in the theoretical isochrones and
removing these degeneracies in the fits would require
photometry reaching the lower main sequence,
a level well below the limits of the data analyzed here. 

In the absence of more precise information,
a metal enrichment history was assumed in all of the models that is
broadly consistent with the above observational constraints.  The
metallicity of the stars was assumed to increase linearly from
$Z=0.00024$ (${\rm [Fe/H]}=-1.9$) to $Z=0.0019$ (${\rm [Fe/H]}=-1.0$)
from 15 Gyr ago to the present. While there is no spread  in
metallicity allowed, the metallicity is assumed to increase
smoothly within each time bin and thus not all of the stars within
each time bin have the same metallicity.  This metal enrichment law
was held constant for all of the solutions described below. The
effects of varying the initial and final metallicities as well as the
enrichment law are described in \citet{wyder01}.  As in
\citet{wyder01}, the isochrones are populated with stars assuming a
power-law initial mass function with exponent $-1.35$ that extends
from 0.1 to 120 $M_{\odot}$.

The extinction $A_V$ was assumed to be constant for all the stars
within each field and the
corresponding reddening was calculated from the Galactic reddening law
of \citet{odonnell94} with $R_V=A_V/E(B-V)=3.1$. In \citet{wyder01}
the extinction was allowed to vary while keeping the distance constant
whereas here both the extinction and distance were varied.

For each field, a set of solutions was generated for various
combinations of extinction and distance. First, a coarse run was
completed with step sizes of 0.1 mag in $A_V$ and $(m-M)_0$. Then I
ran a second set of solutions for each field with smaller step sizes
of 0.05 mag in $(m-M)_0$ and 0.02 mag in $A_V$ with the range of
values restricted to be within $\pm0.2$ mag of the minimum found in
the coarse runs.  The combination of $A_V$ and $(m-M)_0$ with the
smallest value of $\Upsilon$ is then assumed to be the best-fit solution.

Estimating the uncertainties in both the fitted parameters (i.e. $A_V$
and $(m-M)_0$) and the SFRs, is difficult in this case mainly because
none of the solutions provide an acceptable fit to the data in an
absolute sense. According to the discussion in \citet{dolphin02}, the
ideal way to estimate the uncertainty is to determine the average and
variance of $\Upsilon$ from the model itself by assuming that the
number of stars in each model bin follows the Poisson distribution.
Then the uncertainties in all of the SFH parameters can be estimated
by looking at the distribution of parameters for all solutions with
values of $\Upsilon$ with corresponding probabilities greater than
some specified value. Since all of the solutions presented here are
fairly far away from the true best-fit value of $\Upsilon$, this ideal
procedure does not apply.

An alternative method of estimating the uncertainties in the values of
$A_V$, $(m-M)_0$ and SFR is to construct bootstrap samples from the
data \citep[e.g.][]{olsen99}.
For each field, I generated 25 bootstrap versions of the
photometry list by randomly choosing stars from the observed list with
replacement. Then the minimum of $\Upsilon$ was found for each
bootstrap sample using exactly the same procedure as for the original
data. The standard deviations of $A_V$, $(m-M)_0$ and the SFR in each
bin among the solutions for the bootstrap samples are used in the next
section to estimate the statistical uncertainties in the fitted
parameters.

\section{Results and Discussion}

In this section the results of the SFH solutions described in the
previous section are presented.

\subsection{Comparison of model and observed Hess diagrams}

\begin{figure}
\plotone{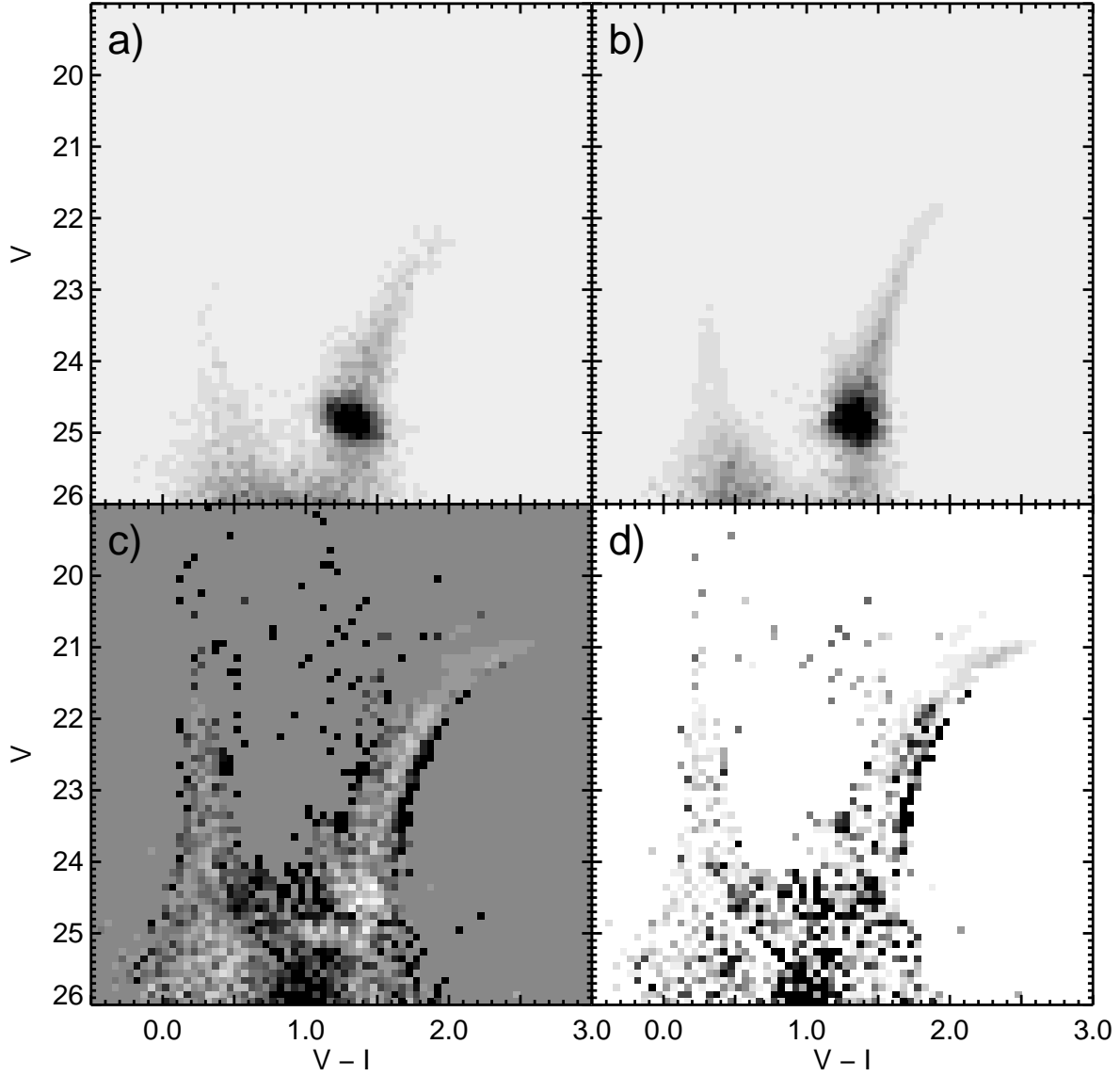}
\caption{Comparison of the observed and model Hess diagrams for the C1
field. Panels (a) and (b) show the observed and best-fit model Hess
diagrams, respectively.  Panel (c) shows the difference between the
observed and model diagrams  divided by the  uncertainty in the model
diagram.  The data in this panel are plotted from $-5$ (white) to $+5$
(black) and thus the black areas are where there are more observed
than model stars. Panel (d) plots the contribution of each Hess
diagram bin to the overall value of $\Upsilon$ as defined in equation
(6).\label{c1hess}}
\end{figure}

\begin{figure}
\plotone{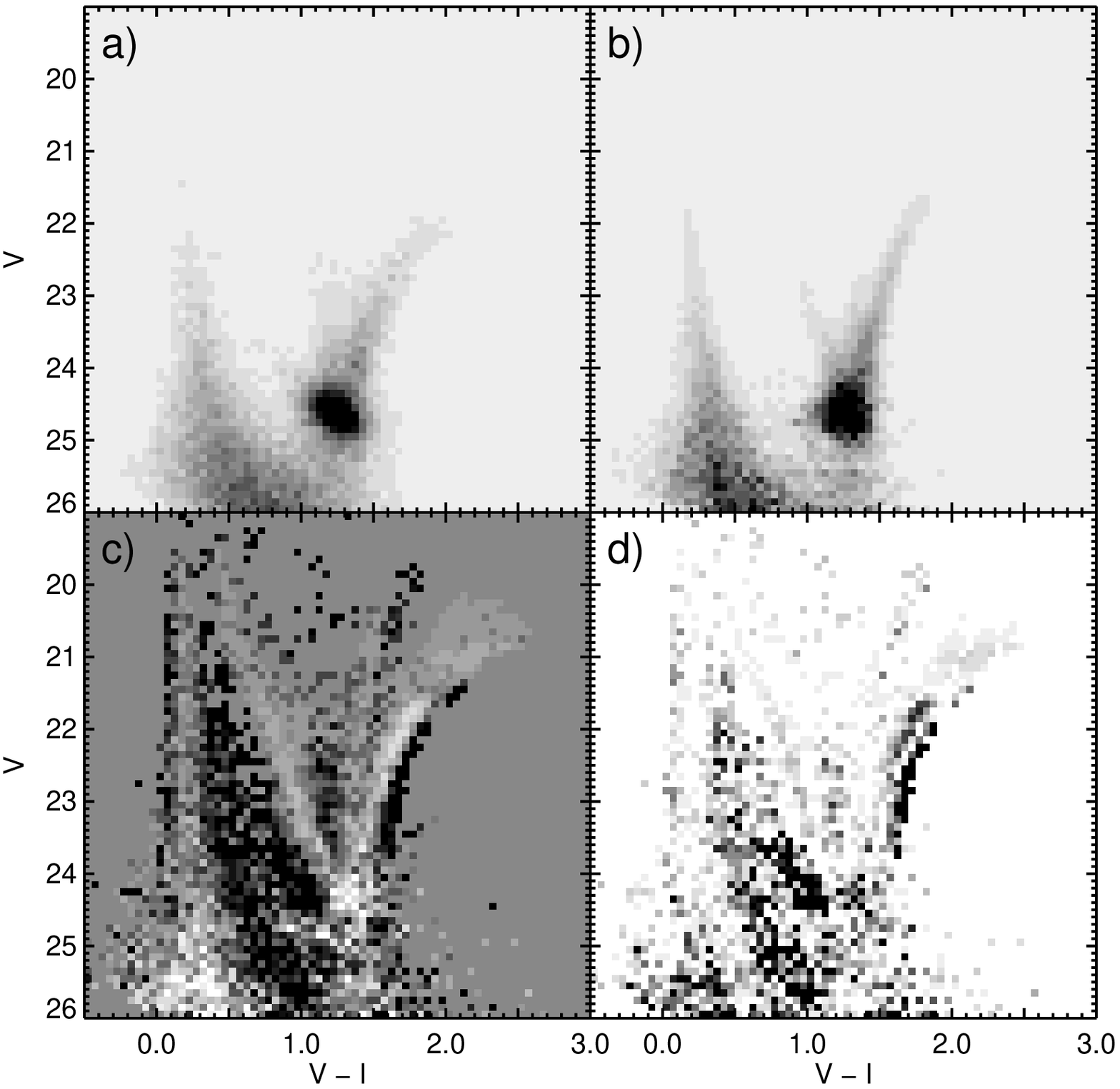}
\caption{Same as Figure \ref{c1hess}, except for the C12 field.\label{c12hess}}
\end{figure}

\begin{figure}
\plotone{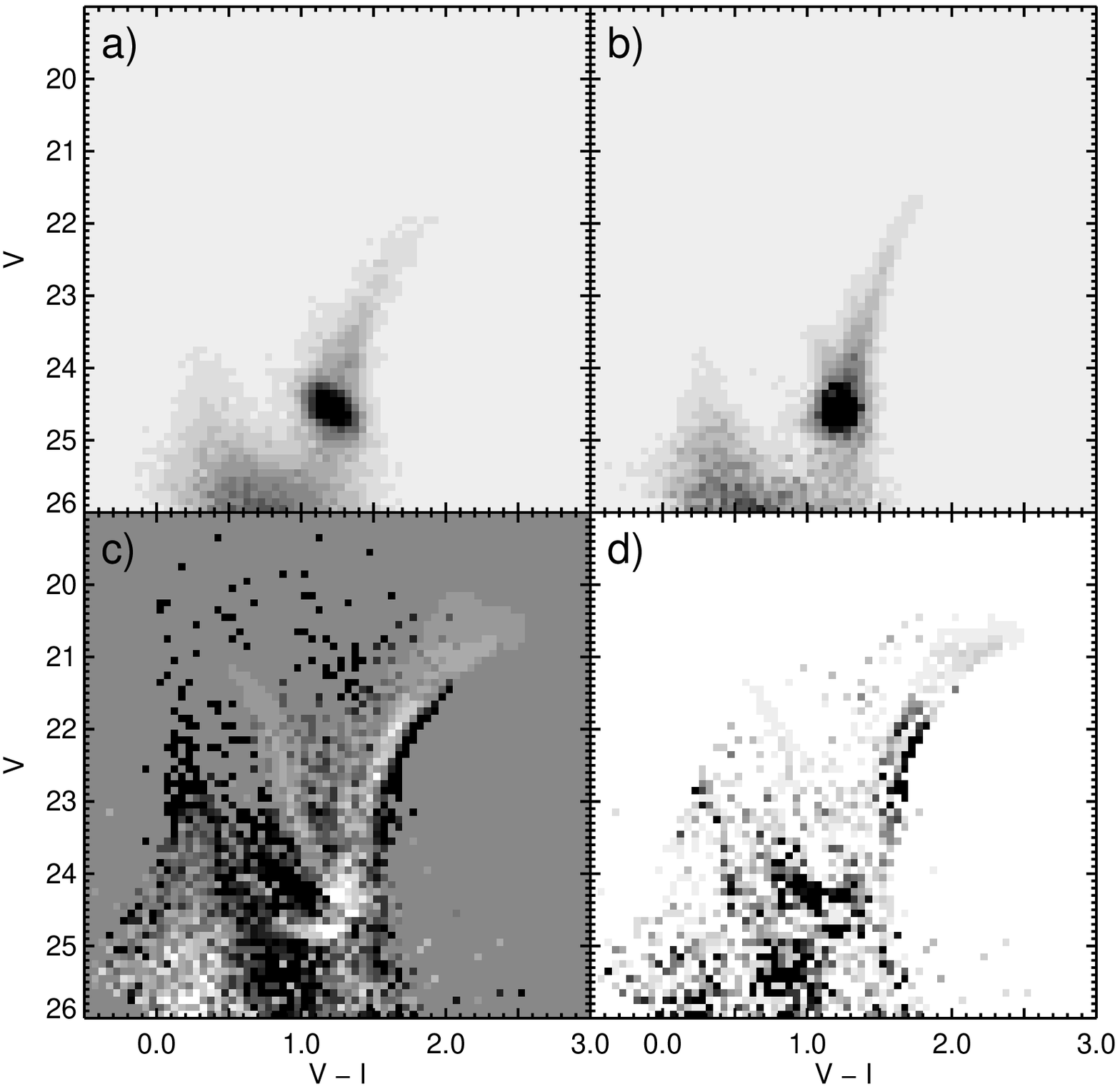}
\caption{Same as Figure \ref{c1hess}, except for the C18 field.\label{c18hess}}
\end{figure}

\begin{figure}
\plotone{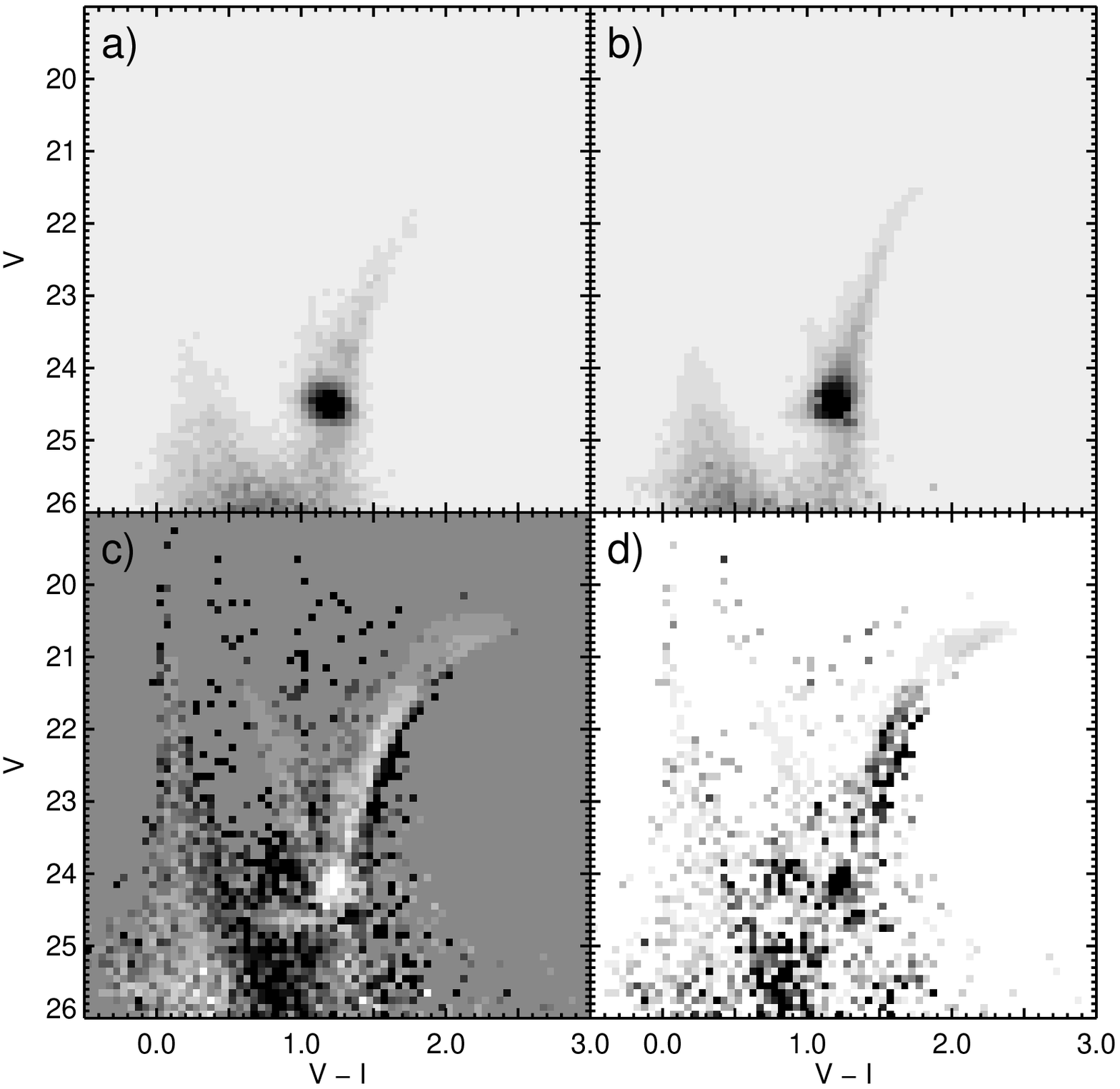}
\caption{Same as Figure \ref{c1hess}, except for the C25 field.\label{c25hess}}
\end{figure}

In Figures \ref{c1hess}$-$\ref{c25hess}, the best-fit model Hess
diagrams are compared with the observed diagrams. In each of the
figures, the observed and best-fit model Hess diagrams are shown in
panels (a) and (b), respectively. Panel (c) shows the
difference between the observed and model diagrams divided by the
expected uncertainty in the model diagram which is assumed to be given
by the square root of the number of model stars in each Hess diagram
bin. The latter diagram is scaled from $-5$ (white) to $+5$ (black).
Finally, a map of the contribution to the overall $\Upsilon$ sum
in equation (6) is shown in panel (d).

Overall, the model Hess diagrams contain the major features that
are in the observed diagrams and reproduce their relative strengths.
However, it is clear from
Figures \ref{c1hess}$-$\ref{c25hess} that the models and observations
do differ in detail.
Except for the outermost C1 field, the plume of blue main sequence stars 
is wider in the observations than in the models. Since the effects
of errors in the photometry are included in the models using the
artificial star test results, the observed width is not due
to errors alone. One explanation for the difference is that each
of the model diagrams assumes a single value for the extinction
and reddening within each field. If a range of
reddenings were allowed in the models within each of the fields,
then the main sequence would widen.

In most of the fields, the models do not reproduce the observed stars
at the red edge of the red giant branch. The shape of the red clump is
different as well between the models and data.  Since the colors of
the red giant branch stars and the red clump morphology are strongly
affected by metallicity, this difference could indicate that the
age-metallicity relation I have adopted is incorrect. On the other
hand, the red clump tilts up and to the left in the observed Hess
diagrams, particularly in the C12 and C18 fields, and is oriented
roughly parallel to the reddening vector. Hence, a range of reddenings
within each field might also account for part of the differences
between the observations and models.

Finally, it is important to note that there are significant
uncertainties in the stellar evolutionary models which may account for
some of the differences between the model and observed Hess diagrams.
For example, \citet{gallart03} compared the Padova isochrones
\citep{girardi00} in the $1-3$ Gyr age range with those from
\citet{yi01} and found significant differences in the
shapes of the RGB, subgiant branch and main sequence turn-off. 
These differences were attributed
to different assumptions made regarding the input physics such as
convective overshoot and the equation of state. In addition, the
methods used to transform luminosities and effective temperatures
into observable magnitudes and colors are different between the two
isochrone sets and lead to differences in the resulting tracks in
the CMD. 

\subsection{Distance}

The distance moduli corresponding to the best-fit SFHs are listed in
Table \ref{besttab} while Table \ref{bstab} gives the results of the
solutions for the bootstrap samples. Columns (2) and (3) of Table
\ref{bstab} list the average and standard deviation, respectively,
calculated from the solutions of the bootstrap samples. The average
values from the bootstrap samples are similar to the best-fit values
from the solutions to the original data, indicating that there is no
bias inherent in the Hess diagram fitting itself. The standard
deviations from the bootstrap fits
have values of $0.03-0.06$ mag and I assume these values as
the statistical error in the distances due to the fitting of the
observed Hess diagrams. This does not include any potential systematic
errors due to the various assumptions made in construction of the
model Hess diagrams.  These best-fit distance moduli of 23.15-23.25
are smaller than the red clump value of $(m-M)_0=23.34\pm0.10$ determined
in \S{3.2} as well as the
distance moduli determined by \citet{gallart96c}
from both Cepheids and the TRGB which
yielded values of $23.49\pm0.08$ and $23.4\pm0.1$, respectively. The
difference between the distances determined here and the
Cepheid, TRGB  and red clump determinations are larger than the random errors
quoted in Table \ref{bstab}. However, there are likely systematic
errors in the distances determined from the CMD-fitting due to the
assumptions made about the metal enrichment as I argue below.

While the distances determined from the SFH solutions are sensitive to
features in the entire CMD, the use of the Poisson fit parameter means
that those bins of the Hess diagram with the largest numbers of stars
will contribute most to the best-fit value of the distance. As is
obvious from the CMDs in Figure \ref{cmds}, the red clump is the most
prominent feature in all of the diagrams and its magnitude has a large
influence on the derived value of the distance.

If I calculate the red clump absolute
magnitudes in the same way as in \S{3.2} except using the same age-metallicity
relation as used in the SFH solutions, I obtain values of
$M_V=+0.301$ and $M_I=-0.517$, leading to a distance modulus of
$(m-M)_0=23.31$ from both the $V$ and $I$ magnitudes.
These values are $\approx 0.1$ mag
larger than the values determined from the fits to the full CMD listed in
Table \ref{besttab} despite adopting the identical age-metallicity
relation and SFR(t).  While the red clump is the most dominant feature
in each of the CMDs and has a large influence in the best-fitting
distance, it is important to note that it is the entire CMD that
determines the distance. Hence, there must be other features in the
CMD which are also affecting these distance
determinations. Furthermore, the SFH is fit to the $V,V-I$ CMD and
thus it is the $V$-band absolute magnitude of the models which directly
determines the distance. As has been pointed out previously
\citep[i.e.][]{girardi01}, the value of $M_V$ for the red clump is
more sensitive to age and metallicity effects than $M_I$.

The difference of $\approx 0.4$ mag between the value of $M_V$ for
${\rm [Fe/H]}=-0.7$ and ${\rm [Fe/H]}=-1.7$ is an estimate of the systematic
effects of the unknown age-metallicity relation on the distance
determined from the fits to the observed CMDs. Thus, the
distance moduli listed in Table \ref{besttab} from the SFH
solutions have systematic
uncertainties of $\pm\sim0.2$ mag due simply to metallicity effects.
Averaging the results for the four fields yields a distance
modulus of $(m-M)_0=23.20\pm0.05(\pm0.2)$ where the first error
is the statistical error from the bootstrap analysis and the
error in parentheses is the systematic error due to our lack of
knowledge about the chemical enrichment history.

These results underscore the caution
that must be taken when attempting to determine the distance from
fitting the CMD. It is important
to understand the systematic effects of the assumptions made regarding
the parameters used in generating the model Hess diagrams.

\subsection{Extinction}

The best-fitting values of the extinction in each field are listed  in
Table \ref{besttab} whereas the results of the bootstrap analysis are
shown in Table \ref{bstab}. As was the case for the distance
determinations, the average values of $A_V$ from the bootstrap samples
are nearly the same as the values from the best-fit to the actual
data, indicating no inherent bias in the value of $A_V$ due to the
fitting procedure alone. The standard deviations for each of the
fields from the bootstrap analysis are $\approx 0.01$ mag.  Estimating
any systematic errors in the extinction due to the assumptions made
about the metallicities and other parameters is more difficult since
the determination of the extinction is sensitive to both the color and
magnitude of features in the CMD. However, one crude estimate can be
made by examining the spread in the red clump magnitudes $V_{0,m}$
listed in Table \ref{rclftab}. Since these values have been corrected
for the extinction using the values from fitting the CMD, they should
yield the same number if any differences in stellar
populations among the four fields are neglected. 
Half of the full spread of values
for $V_{0,m}$ is 0.05 mag which I will assume as the uncertainty in
the extinction values listed in Table \ref{besttab}.

There is a trend of increasing extinction going from fields C25 to C1,
or equivalently, going from east to west across the minor axis of the
galaxy. While this direction also corresponds to decreasing distance
from the Galactic plane, the difference of $\approx 0.4$ mag among
the four fields across only $\approx 10\arcmin$ would be difficult to
explain as variations in the Galactic foreground extinction.
The extinction determined for the C25 field of $A_V=0.84$ is
comparable to the foreground Galactic extinction to NGC 6822 of
$A_V\approx0.7$ as determined from the \citet{schlegel98} reddening
maps. The higher extinction values in the other three fields
imply the existence of additional extinction internal to NGC 6822
itself. From observations of OB stars in NGC 6822, \citet{massey95}
found a similar increase in the reddening from
$E(B-V)=0.26$ ($A_V=0.81$) in the outer regions to $E(B-V)=0.45$
($A_V=1.4$) in the bar. A similar trend was found for the four other
{\it HST} fields presented in \citet{wyder01} where the extinction in
the bar was determined to be $A_V\approx1.0$ and smaller in the outer
regions. The one exception to these trends is the C1 field where the
best fit extinction is $A_V=1.2$ despite the fact that this is the
field located the farthest from the bar.

\begin{figure}
\plotone{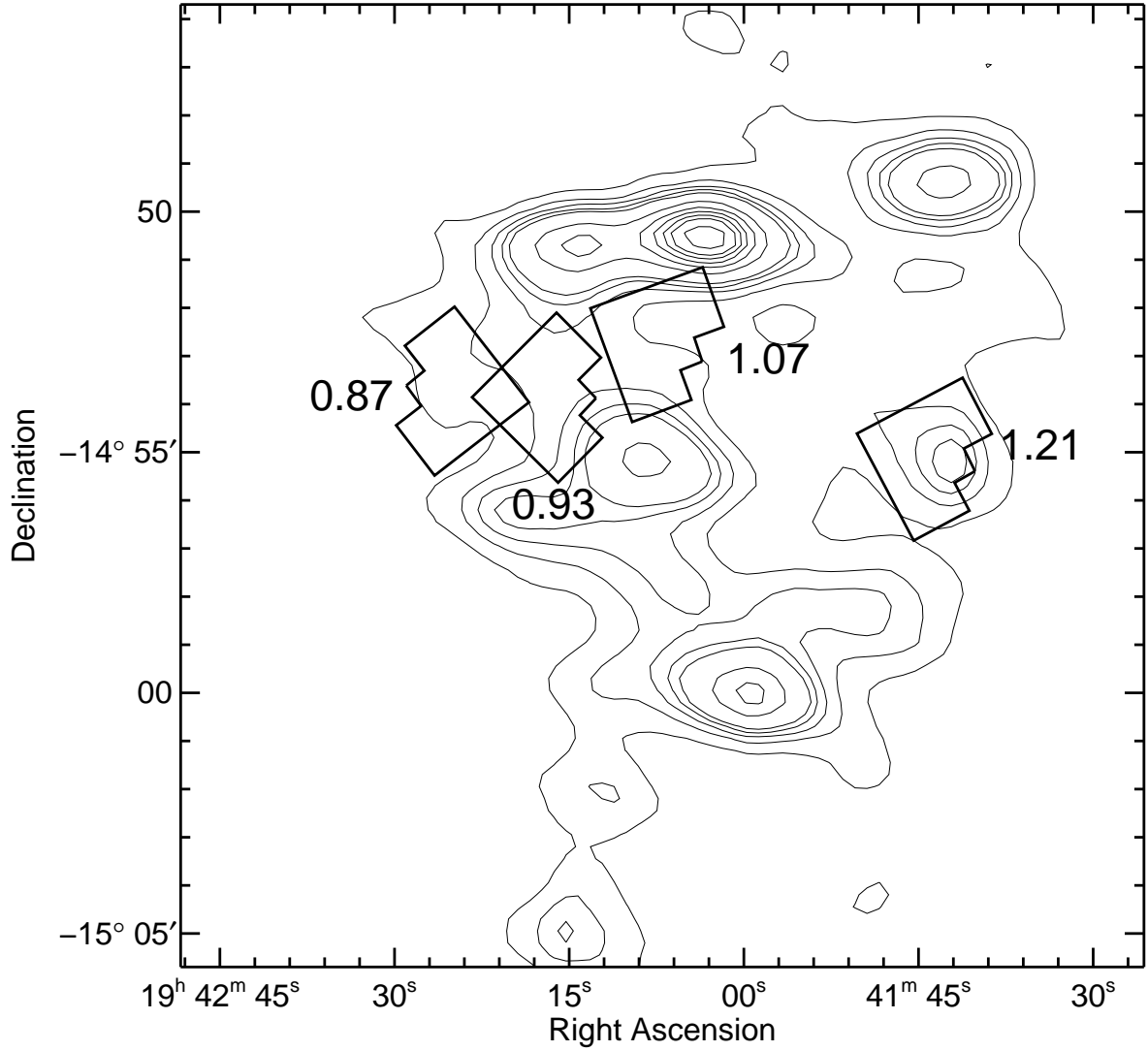}
\caption{Location of the four WFPC2 fields overlayed on the $60\micron$
emission measured by the IRAS satellite \citep{rice93}. The coordinates
are epoch
1950 and the value of the extinction $A_V$ determined from the CMD fitting
is shown next to each field. The contour levels corresponding to surface
brightnesses of 1, 2, 3, 4, 5, 10, 15, 20, 25, 30 and 40 MJy sr$^{-1}$.
\label{iras}}
\end{figure}

The dust responsible for the extinction internal to NGC 6822 is
expected to emit at far infrared wavelengths.  A contour map of the
$60\micron$ emission in NGC 6822 from \citet{rice93} as measured by
the IRAS satellite is shown in Figure \ref{iras} where the locations
of the four {\it HST} fields analyzed here are indicated. As in
\citet{wyder01}, the extinction values derived from fitting the CMD
are roughly correlated with the FIR emission. In fact the C1 field
which has the largest value of the extinction is coincident with a
local peak in the $60\micron$ emission. This FIR peak contains the
\ion{H}{2} regions K$\alpha$, K$\beta$ and HK 1 \citep{hodge88} and it
would be reasonable to expect that the dust responsible for the
extinction in the C1 field and this peak in the FIR emission is
associated with these star formation regions.  In general there is a
good correspondence between the locations of \ion{H}{2} regions and OB
associations in NGC 6822 and peaks in the FIR emission although there
are substantial spatial variations in the ratio of FIR to H$\alpha$
luminosity \citep{gallagher91,israel96}. The amount of FIR emission is
expected to be proportional to both the amount of dust and the
strength of the UV radiation field.  However, star formation regions
are also those areas where increased densities of gas and dust are
expected.  The correlation between the FIR surface brightness and the
extinction inferred from the CMDs suggests that the varying FIR emission is
indeed tracing the distribution of dust in NGC 6822 rather than solely
being due to variations in the interstellar radiation field.

\subsection{Star Formation Histories}

\begin{figure}
\plotone{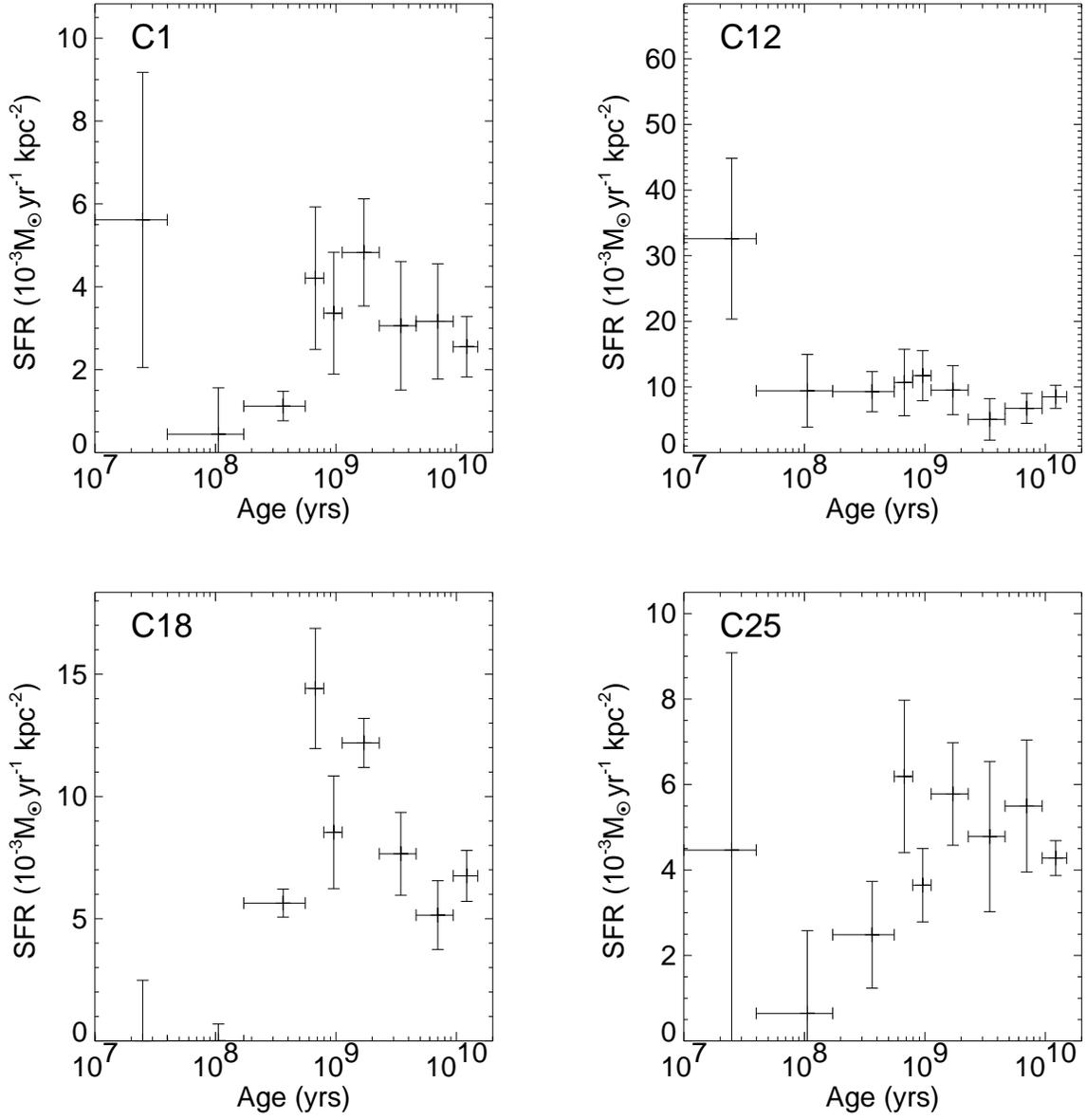}
\caption{The star formation rate per area as a function of age from
the present for all four pointings in NGC 6822.\label{sfh1}}
\end{figure}

The SFRs per area averaged over the field-of-view for each of the four
fields are plotted in Figure \ref{sfh1}. The error bars in age denote
the limits of the time bins chosen for the SFH solutions.  The error
bars in the SFRs in each time bin are computed from the fits to the 25
bootstrap samples of each CMD. These uncertainties are calculated as
the standard deviations of the SFRs in each bin.  It is important to
emphasize that the SFRs in each of the bins are not independent since
the total number of stars for any model is conserved. Therefore, a
decrease in the SFR in one time bin is accompanied by an increase
in the SFR in another bin.

For the older ages ($>1$ Gyr), the shapes of the SFHs among the four fields
are similar although the overall normalization varies due to the
different numbers of stars within each field. The SFRs are fairly constant
or perhaps slightly increasing with time in some of the fields.
For the C1, C18 and C25 fields, the SFR declines at an age of $\sim 600$ Myr
by a factor of $\sim 2-4$ while for the C12 field centered on the
bar, the SFR remains roughly constant four ages less than 600 Myr with
an increase for the youngest time bin. Despite the drop in star formation
seen in the three other fields, there are some stars present in these
fields with ages as young as a few 10s of Myr old. Thus, the outer 
regions sampled here do not represent a pure population II halo as many
larger galaxies possess. 

As shown by \citet{kennicutt98}, the relationship between the global
average SFR per area and the surface density of atomic plus molecular
gas in galaxies is well represented by the following relation
\begin{equation}
\Sigma_{SFR} = (2.5\pm0.7) \times 10^{-4} 
\left(\frac{\Sigma_{gas}}{1~{\rm M_{\sun} pc^{-2}}}\right)^{1.4\pm0.15}
{\rm M_{\sun}~yr^{-1}~kpc^{-2}}
\end{equation}
This global relationship can be used to compare the SFRs determined
here from the CMD-fitting with the SFRs implied by the star formation
law in equation (7). Using the position angle, center and inclination
derived from the  \ion{H}{1} observations of \citet{weldrake02}, the
four {\it HST} fields lie at galactocentric radii of $\sim
350-500\arcsec$. At these radii, the azimuthally averaged \ion{H}{1}
surface density is $\Sigma_{HI} \approx 4~{\rm M_{\odot} pc^{-2}}$
\citep{weldrake02}. Of the galaxies used in the \citet{kennicutt98}
sample, the \ion{H}{1} gas accounts for one-half the total gas mass on
average although there is a large dispersion in the atomic gas
fraction in the sample. Assuming that one-half of the gas mass in NGC
6822 is \ion{H}{1}, the total gas surface density at the radii sampled
by the {\it HST} fields is $\Sigma_{gas} \approx 8$ M$_{\sun}$
pc$^{-2}$. According to equation (7), this corresponds to a SFR
surface density of $\Sigma_{SFR}=(4.6\pm1.3) \times 10^{-3}~{\rm
M_{\sun}~yr^{-1}~kpc^{-2}}$ where the uncertainty is derived from the
errors in the parameters in equation (7). This SFR surface density is
comparable to that observed for the average old ($>1$ Gyr) SFRs in all
the fields. This approximate agreement between the SFRs from the
global SFR law and the SFRs calculated from the CMDs means that it
would have been possible for the gas in NGC 6822 to be distributed
many Gyr ago as it is today and be able to produce the SFRs that have been
determined for the older stars. Therefore, there is no need for a
large scale redistribution of gas or stars to account for the SFRs
shown in Figure \ref{sfh1}.

With the exception of the C12 field centered on the bar, the recent
SFRs are smaller than the Schmidt law value. However, star formation
in most galaxies tends to be quite patchy and occurs in associations
of  various sizes that gradually dissolve over time. In particular,
the C12 field contains part of OB associations 9 and 11 from the list
in \citet{hodge77}. The C25 field contains part of OB association 15
and the remaining two fields contain none of the associations from
\citet{hodge77}. Nevertheless, the recent star formation is not
entirely confined to the OB associations since there are bright main
sequence stars in each pointing scattered across the field-of-view.
Since each field samples a relatively small area, the precise values
of the recent SFRs in each field are sensitive to this irregular
spatial distribution of young stars. As a result, the SFR surface
densities in each field may differ from the azimuthally averaged
value. Furthermore, the gas density is also somewhat clumpy which
would cause the local gas density to differ from the azimuthally
averaged value as well.

In their ground-based data covering a much larger area of NGC 6822,
\citet{gallart96b} found that the SFR $100-200$ Myr ago
increased by a factor of 2 in
the outer regions and in the center of the bar and
increased by a factor of $4-6$ at the ends of the bar. \citet{deblok00}
detected a $2.0 \times 1.4$ kpc$^{2}$ hole in the \ion{H}{1} gas lying
to the east of the bar that would have an age of roughly 100 Myr based
upon its size. One possible explanation for the increase in SFR and
the creation of the \ion{H}{1} hole $\sim 100$ Myr ago is an
interaction between the main body of the galaxy and the \ion{H}{1}
cloud to the NW of the bar \citep{deblok00}. It would be reasonable to
assume that similar variations in the SFR have occurred throughout the
history of NGC 6822 due to similar relatively minor interactions.
Since the resolution in age for the SFH solutions described here
decreases quickly with increasing age, it would be possible for short
term star formation bursts to remain undetected in the current
data. However, the overall picture that can be drawn from the SFHs in
Figure \ref{sfh1} is that NGC 6822 has been forming stars at a more or less
constant rate over its lifetime with no evidence for large long term
variations in the SFR. This picture is consistent with the relatively
isolated position that NGC 6822 occupies in the Local Group and the
corresponding lack of major interactions that it has likely experienced.

\acknowledgments
I am grateful to Andrew Dolphin for providing me with this HSTPHOT
photometry programs as well as for his generous help in assisting me
to understand his star formation history analysis methods. I am also
grateful to Paul Hodge for allowing me to work on this data and his
encouragement and guidance.  Support for proposal 8314 was provided by
NASA through a grant from the Space Telescope Science Institute, which
is operated by the Association of Universities for Research in
Astronomy, Inc., under NASA contract NAS 5-26555.

\clearpage

\clearpage

\begin{deluxetable}{lcccc}
\tablecaption{Summary of observations\label{obs_log}} \tablewidth{0pt}
\tablehead{ \colhead{Field} & \colhead{Dataset} & \colhead{Filter} &
\colhead{Exp. time (sec)} & \colhead{Date of obs.}  }
\startdata 
C1   & u5ch0103r & F555W & 800 & 2000 Jun 3 \\  
   & u5ch0104r & F555W & 400 & 2000 Jun 3 \\
   & u5ch0105r & F814W & 600 & 2000 Jun 3 \\
   & u5ch0106r & F814W & 600 & 2000 Jun 3 \\
C12  & u5ch0203r & F555W & 800 & 2000 Jun 3 \\
  & u5ch0204r & F555W & 400 & 2000 Jun 4 \\
  & u5ch0205r & F814W & 600 & 2000 Jun 4 \\
  & u5ch0206r & F814W & 600 & 2000 Jun 4 \\
C18  & u5ch0303r & F555W & 800 & 2000 Jun 4 \\
  & u5ch0304r & F555W & 400 & 2000 Jun 4 \\
  & u5ch0305r & F814W & 600 & 2000 Jun 4 \\
  & u5ch0306r & F814W & 600 & 2000 Jun 4 \\
C25  & u5ch0403r & F555W & 800 & 1999 Sep 24 \\
  & u5ch0404r & F555W & 400 & 1999 Sep 24 \\
  & u5ch0405r & F814W & 600 & 1999 Sep 24 \\
  & u5ch0406r & F814W & 600 & 1999 Sep 24 \\
\enddata
\end{deluxetable}

\begin{deluxetable}{lccccc}
\tablecaption{Tip of the red giant branch distance to NGC 6822\label{rgbtab}}
\tablewidth{0pt}
\tablehead{\colhead{Field} & \colhead{$I_{TRGB}$} & \colhead{$(V-I)_{-3.5}$} &
\colhead{$(V-I)_{TRGB}$} & \colhead{$(m-M)_0$}}
\startdata
C1  & $20.30\pm0.30$ & 1.77 & 1.94 & $23.56\pm0.30$ \\
C12 & $20.00\pm0.10$ & 1.77 & 1.90 & $23.38\pm0.10$ \\
C18 & $19.90\pm0.10$ & 1.75 & 1.88 & $23.36\pm0.10$ \\
C25 & $19.95\pm0.15$ & 1.71 & 1.87 & $23.47\pm0.15$ \\
\enddata
\end{deluxetable}

\begin{deluxetable}{lcccccc}
\tablecaption{Gaussian fits to the red clump magnitude distribution
\label{rclftab}}
\tablewidth{0pt}
\tablehead{\colhead{Field} & \colhead{$N_{RC,V}$} & \colhead{$V_{0,m}$} &
\colhead{$\sigma_{RC,V}$} & \colhead{$N_{RC.I}$} & \colhead{$I_{0,m}$} &
\colhead{$\sigma_{RC,I}$}}
\startdata
C1  & 122 & 23.63 & 0.16 & 140 & 22.81 & 0.16 \\
C12 & 273 & 23.56 & 0.19 & 306 & 22.76 & 0.18 \\
C18 & 258 & 23.60 & 0.17 & 318 & 22.79 & 0.17 \\
C25 & 178 & 23.66 & 0.15 & 201 & 22.82 & 0.16 \\
\enddata
\end{deluxetable}

\begin{deluxetable}{lcc}
\tablecaption{Best fit values of $A_V$ and $(m-M)_0$\label{besttab}}
\tablewidth{0pt}
\tablehead{\colhead{Field} & \colhead{$(m-M)_0$} & \colhead{$A_V$}}
\startdata
C1 & 23.20 & 1.20 \\
C12 & 23.15 & 1.06 \\
C18 & 23.25 & 0.94 \\
C25 & 23.20 & 0.84 \\
\enddata
\end{deluxetable}

\begin{deluxetable}{lcccc}
\tablecaption{Results of bootstrap solutions\label{bstab}}
\tablewidth{0pt}
\tablehead{\colhead{Field} & \colhead{$\langle(m-M)_0\rangle$} &
\colhead{$\sigma(m-M)_0$} & \colhead{$\langle A_V \rangle$} &
\colhead{$\sigma(A_V)$}}
\startdata
C1  & 23.208 & 0.059 & 1.208 & 0.017 \\
C12 & 23.150 & 0.050 & 1.065 & 0.010 \\
C18 & 23.238 & 0.033 & 0.950 & 0.010 \\
C25 & 23.232 & 0.045 & 0.839 & 0.011 \\
\enddata
\end{deluxetable}

\end{document}